\documentclass[letter, twocolumn]{jpsj3}
\usepackage{txfonts}

\newcommand{\Fig}{Fig. }
\newcommand{\Tab}{Table }
\newcommand{\this}{letter}
\usepackage{color}

\title{Detection of phase transition via convolutional neural networks}

\author{Akinori Tanaka$^1$\thanks{akinori.tanaka@riken.jp} and Akio Tomiya$^2$\thanks{akio.tomiya@mail.ccnu.edu.cn}}
\inst{$^1$Interdisciplinary Mathematical and Computational Collaboration Team, 
RIKEN, Wako 351-0198, Japan \\
$^2$Key Laboratory of Quark \& Lepton Physics (MOE) and Institute of Particle Physics, Central China Normal University, Wuhan 430079, China} %\\

\abst{
A Convolutional Neural Network (CNN) is designed to study correlation between the temperature and the spin configuration of the 2 dimensional Ising model.  Our CNN is able to find the characteristic feature of the phase transition without prior knowledge.  Also a novel order parameter on the basis of the CNN is introduced to identify the location of the critical temperature; the result is found to be consistent with the exact value.
}

\begin{document}
\maketitle

%\section{Introduction} \label{Introduction} 
%\vspace{-0.2cm}
%\vspace{-0.2cm}
Studies of phase transition are connected to various areas among theoretical/experimental physics \cite{wilson1974renormalization, polchinski1992effective, intriligator1996lectures, pasechnik2016phenomenological, hohenberg2015introduction, chen2005bcs, kramer2005random}.
Calculating order parameters is one of the conventional ways to define phases and phase transitions.
However, some phases like topological phases \cite{wen1990topological} do not have any clear order parameters.
Even if there are certain theoretical order parameters like entanglement entropy \cite{kitaev2006topological, levin2006detecting}, they are difficult to measure in experiments.

Machine learning (ML) techniques are useful to resolve this undesirable situation.
In fact, ML techniques have been already applied to various problems in theoretical physics:
finding approximate potential surface \cite{behler2007generalized}, 
a study of transition in glassy liquids \cite{schoenholz2016structural},
solving mean-field equations \cite{arsenault2015machine} and quantum many-body systems \cite{arsenault2014machine, carleo2016solving},
a study of topological phases \cite{deng2016exact}.

Especially, ML techniques based on convolutional neural network (CNN) have been developing since the recent groundbreaking record \cite{krizhevsky2012imagenet} in ImageNet Large Scale Visual Recognition Challenge 2012 (ILSVRC2012) \cite{ILSVRC15}, and it is applied to investigate phases of matters with great successes on classifications of phases in 2D systems \cite{carrasquilla2016machine, broecker2016machine, ohtsuki2016deep2D} and 3D systems \cite{ch2016machine,  ohtsuki2016deep3D}.
It is even possible to draw phase diagrams \cite{ohtsuki2016deep3D}.

In these previous works, however, one needs some informations of the \textit{answers} for the problems {\it a priori}.
For example, to classify phases of a system, the training process requires the values of critical temperatures or the location of phase boundaries.
This fact prevents applications of the ML techniques to unknown systems so far.

%\vspace{-1.0cm}

The learning process without any answers is called {\it unsupervised learning}.
Indeed, there are known results on detecting the phase transitions based on typical unsupervised learning architectures called autoencoder which is equivalent to principal component analysis \cite{wang2016discovering} and its variant called variational autoencoder \cite{wetzel2017unsupervised}. 
These architectures encode informations of given samples to lower dimensional vectors, and it is pointed out that such encoding process is similar to encoding physical state informations to order parameters of the systems.
However, it is not evident whether the latent variables provide the critical temperature.

We propose a novel but simple prescription to estimate the critical temperature of the system via neural network (NN) based on ML techniques
without {\it a priori}~ knowledge of the order parameter.
Throughout this \this, we focus on the ferromagnetic 2D Ising model on a square lattice mainly based on the following three reasons.
First, this system is one of the simplest solvable systems \cite{onsager1944crystal, nambu1995note} which has a phase transition, and it is easy to check the validity of our method.
Second, this system can be generalized to other classical spin systems like Potts model, XY model and Heisenberg model, so our method can be applied to these other systems straightforwardly.
%\cite{wilson1974renormalization, belavin1984infinite, plechko2005fermions}, and 
Third, this model is a good benchmark for new computational methods \cite{mehta2014exact, aoki2009domain}.

\begin{figure}[t]
\centering
\includegraphics[width=7.3truecm,clip]{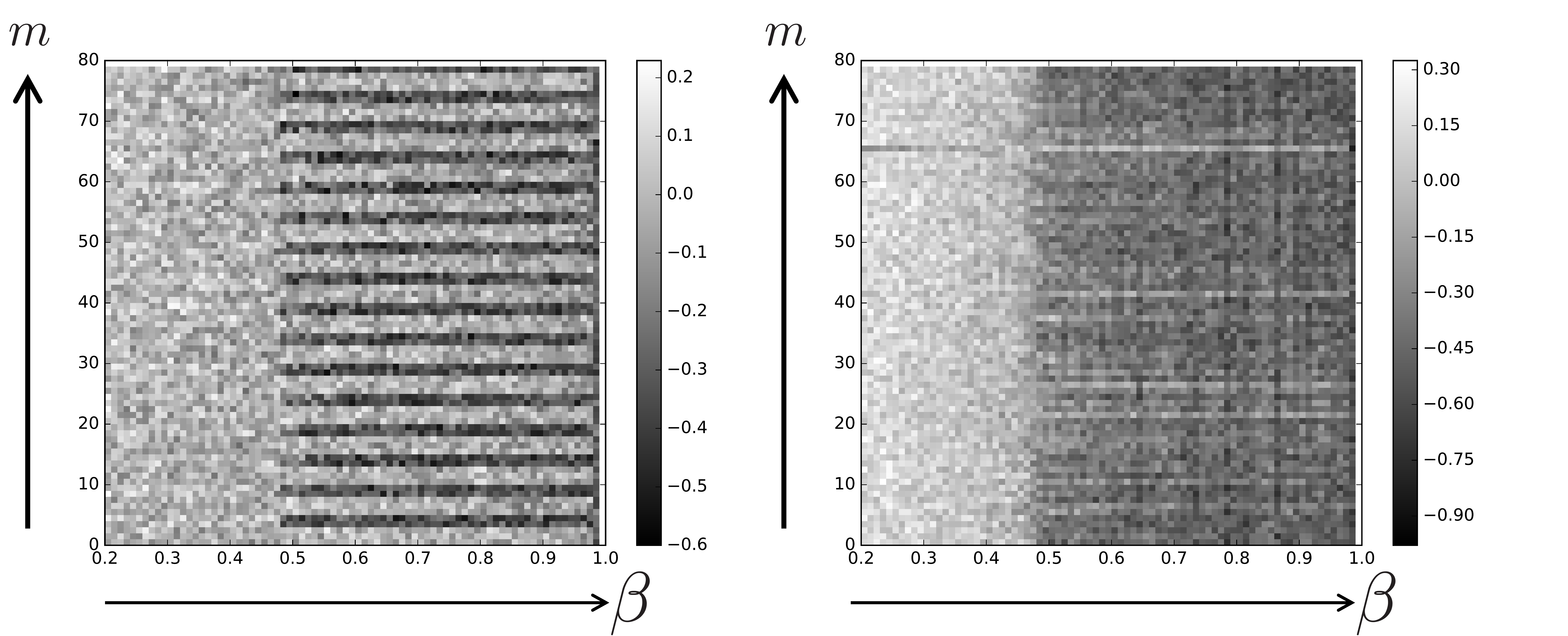} 
\caption{
{`(Color online)'
Plots of weight matrix components in
convolutional neural network (left) and
fully connected neural network (right).
%Components around $0$ are gray colored.
Horizontal axis corresponds to the inverse temperature $\beta$.
Vertical axis $m$ corresponds to components connected to hidden nodes in NN. %One can find \Fig \ref{fig:nn} shows where the $W$ lives in the network.
}
\label{fig:intro}

\vspace{-.5cm}
}
\end{figure}

Using {\tt{TensorFlow (r0.9)}} \cite{abadi2016tensorflow}, we have implemented a neural network to study the correlation between spin configurations and discretized inverse temperatures.
%}.
We find that NNs are able to capture features of phase transition, even for simpler fully-connected (FC) NN, in the weight $W$ without any information of order parameters nor critical temperature.
\Fig \ref{fig:intro} reminds us of the discovery of the ``cat cell" in the literature \cite{le2013building} in which the model recognizes images of cats without having explicitly learned what a cat is.

We examine the boundary structure in \Fig \ref{fig:intro} by defining order parameter, and estimate the inverse critical temperature by fitting distribution of the order parameter (\Tab  \ref{tab:intro}).
\\
\vspace{-.3cm}
 \begin{table}[t]
\begin{tabular}{c|c|c} 
System size& $\beta_c$ (CNN) &$\beta_c$ (FC) \\\hline
8$\times$8 & 0.478915 & 0.462494\\
16$\times$16 &  0.448562& 0.433915\\
32$\times$32 & 0.451887& 0.415596\\
\hline \hline
$L\to\infty$ & 
\multicolumn{2}{c}{$\beta_c^\text{Exact}\sim$ 0.440686}
\end{tabular}
\caption{
Critical temperature of 2D Ising model on the square lattice
extracted from CNN (3 from top). %Bottom column is obtained by the exact solution \cite{onsager1944crystal, nambu1995note}.
$L\to \infty$ stands for thermodynamic limit, and its value is exact one, $\beta_c^\text{Exact}= \frac{1}{2}\log(\sqrt{2}+1) $
\label{tab:intro}

\vspace{-.5cm}
}
\end{table}

\if
This \this \ is organized as follows.
In section \ref{Our model}, we explain our NN model.
In section \ref{Result}, we show experimental results.
In section \ref{Order parameter based on CNN}, we define a new order parameter via the NN and explain how to obtain $\beta_c$ in \Tab \ref{tab:intro}.
Section \ref{Discussion} is devoted to discussion.
\fi

\begin{figure}[t]
\centering
%\vspace{.3cm}
\includegraphics[width=8truecm,clip]{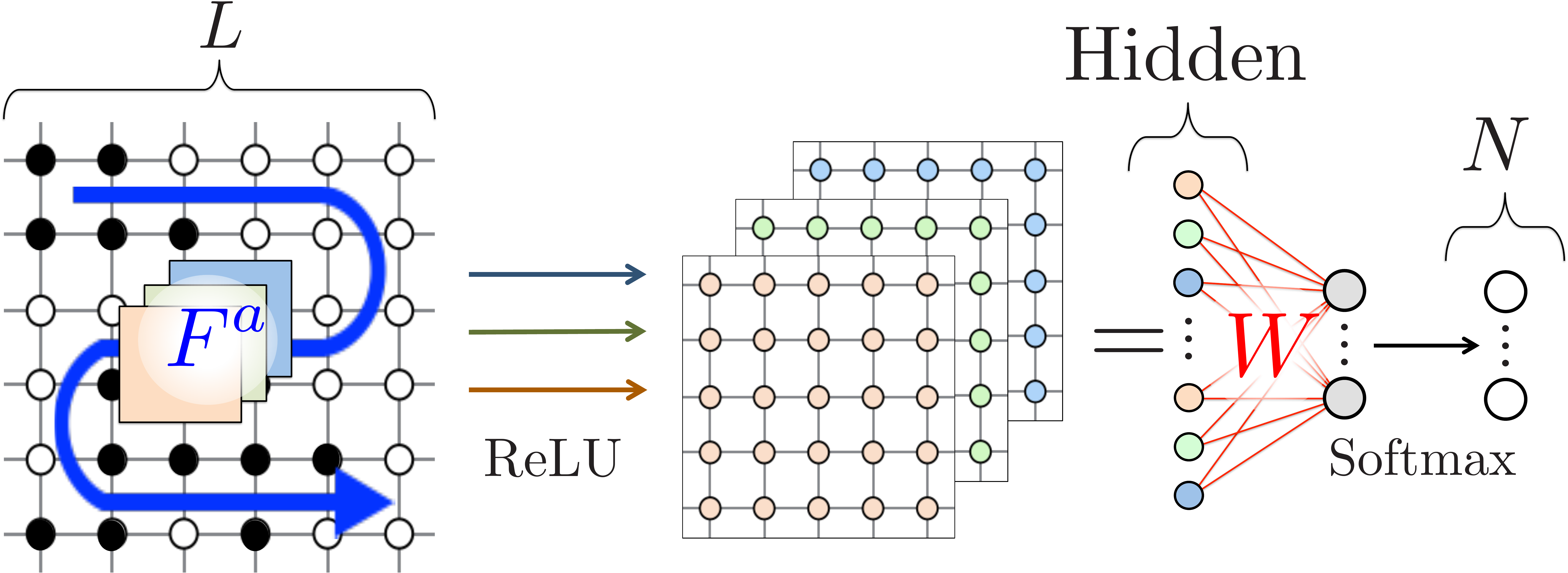} 
\caption{
`(Color online)'
A schematic explanation for the network \eqref{model}. In this figure, we take 3 filters $F^a$ ($a = 1, 2, 3$).
An Ising spin configuration $\{ \sigma_{xy} \}$ generated by MCMC is passed to 3 hidden lattices via the convolution process + ReLU activation.
After making them flatten, they are connected to $N$ nodes via fully-connected weight $W$. In the end, it returns $\beta^\text{CNN}_I$ with the softmax activation.
%The detail of each operation can be found in \eqref{step1} or \eqref{step2}.
\label{fig:nn} 

\vspace{-.8cm}
}
\end{figure}

%\newpage
%\section{Preliminary}\label{Our model}
%
First, we explain the details of our NNs
If the reader is not familiar with machine learning based on NN, we suggest reading literatures reviewing it, e.g. \cite{kolanoski1996application, lecun2015deep}.
Our NN model is designed to solve classification problems.
It is constructed from a convolution layer and a fully connected layer, so it can be regarded as the simplest model of a CNN (\Fig \ref{fig:nn}).
%Usually, a CNN is defined by not only with the convolution layer, but also a \textit{pooling} layer and a \textit{normalization} procedure. In our model, however, we do not introduce these layers for simplicity.
%\footnote{}.
%
The whole definition is as follows.
\begin{align}
\hspace{-.5cm}
\left[ \begin{array}{l}
\mathcal{I}=
\Big\{ \{ \sigma_{xy} \} \Big| \text{ Ising config on }L\times L \text{ lattice.} \Big\}
%\notag 
\\
 \downarrow
\
\left\{ \begin{array}{ll}
\text{Convolution}_{[N_f^2 \text{-filter, } (s,s) \text{-stride, }C \text{-channels}]} & \\
\text{ReLU activation} & \\
\text{Flatten} & \\
\end{array} \right.
%\notag 
\\
\mathbb{R}^{\text{Hidden } = \ L^2 / s^2 \times C }
%\notag 
\\
\downarrow
\ 
\left\{ \begin{array}{ll}
\text{Fully connected} & \\
\text{Softmax} & \\
\end{array} \right.
%\notag 
\\
\text{[0,1]}^{ N  }
 = \mathcal{O}
\end{array} \right]
\label{model}
\end{align}
%
%Each step is defined as follows.
The first transformation in \eqref{model} is defined by {\it convolution} including training parameters $F_{ij}^a$ called \textit{filters}, rectified linear activation called {\it ReLU}, and flatten procedure:
\begin{align}
\hspace{-.3cm}
%&\underline{\text{Convolution + ReLU activation + Flatten:}}
%\notag \\
\left. \begin{array}{l}
\{ \sigma_{xy} \}
\\
\overset{\text{conv}}{\to}
\sum_{i, j = 1}^{N_f}  \sigma_{(s X+i)(s Y+j)} F_{ij}^a = \Sigma_{XY}^a 
\\
\overset{\text{ReLU}}{\to}
\text{max}(0, \Sigma_{XY}^a) = u_{XY}^a
\\
\overset{\text{flatten}}{\to}
\vec{u} = [ u_{11}^1, u_{11}^2, \dots u_{11}^{C}, u_{21}^1, u_{21}^2, \dots, u_{21}^C, \dots %u_{(L^2/s^2)1}^1, u_{(L^2/s^2)1}^2, \dots, u_{(L^2/s^2)1}^C
%\notag \\
%& \qquad
%u_{12}^1, u_{12}^2, \dots u_{12}^{C}, u_{22}^1, u_{22}^2, \dots, u_{22}^C, \dots %u_{(L^2/s^2)1}^1, u_{(L^2/s^2)1}^2, \dots, u_{(L^2/s^2)1}^C
]
=[u_m]
.
\end{array} \right.
&
%\quad 
\label{step1}
\end{align}
The second transformation is defined by fully-connected layer including training parameters $W_I ^m$ called \textit{weights}, and softmax activation:
\begin{align}
%\vspace{.5cm}
%&\underline{\text{Fully connected + Softmax:}}
%\notag \\
& %\qquad
\left. \begin{array}{l}
[u_{m} ]
\\
\overset{\text{fully-connected}}{\to}
\sum_{m=1}^{L^2/s^2\times C} W_I ^m u_m = z_I
\\
\overset{\text{Softmax}}{\to}
\frac{e^{z_I}}{ \sum_{J = 1}^{N } e^{z_J}  }
=
\beta_I^{\text{CNN}}
\end{array} \right.
.
\label{step2}
\end{align}
We classify configurations into $N$ classes labeled by $I$ in the final step.
This $N$ is related to the inverse temperature $\beta$ through \eqref{temp_class}.
Because of the following facts, $\beta_I^\text{CNN} \in [0, 1]$ and $\sum_{I} \beta_I^\text{CNN} = 1$, we can interpret $\beta_I^\text{CNN}$ as a probability for classifying given state $\{\sigma_{xy} \}$ to $I$-th class in a classification problem.
%%%
In total, we have two types of parameters:
\begin{align}
\hspace{-.1cm}
\left. \begin{array}{ll}
F_{ij}^{a} \text{ in convolution,}& 
W^m_{I} \text{ in fully connected layer.} 
\\
\left[ \begin{array}{l}
a = 1, \dots, C  \\
i,j = 1, \dots, N_f  \\
\end{array} \right]
& 
\left[ \begin{array}{l}
m = 1, \dots, L^2/s^2 \times C  \\
I = 1, \dots N  \\
\end{array} \right]
\end{array} \right.
\label{weights}
\end{align}
In later experiment, these parameters will be updated.
In the first trial, we take NN without convolution \eqref{modelFC}.
In this case, the parameters $F$ are not the filters in \eqref{weights} but weights.

\vspace{.7cm}
%\subsection{Preparation of training data for Ising model}\label{prep}
% 
We need training set for optimizing the above parameters \eqref{weights} in CNN.
%Our training set is %not the one in\eqref{pre_training_set}, but the following one:
We call it $\mathcal{T}_L$ where $L$ indicates the size of the square lattice.
The definition is
\begin{align}
%\hspace{-.4cm}
\mathcal{T}_{L}=
\Bigg\{ 
\Big( \{ \sigma_{xy}^{(n)} \}, \vec{\beta}_n \Big) 
\Big|
\frac{1}{\beta_n} = T_\text{min} +  n \delta
\Bigg\}_{n = 0, \dots, (N_\text{conf}-1) }
\hspace{-.4cm}
,
\label{training_set}
\end{align}
where $\{ \sigma_{xy}^{(n)} \} $ is the generated configuration under the Ising Hamiltonian on the square lattice
\begin{align}
H = - \sum_{x, y} \sigma_{xy} \Big(\sigma_{(x+1),y} + \sigma_{x,(y+1)} \Big),
\label{hamiltonian}
\end{align}
and inverse temperature $\beta_n$ using the Metropolis method.

$T_\text{min (max)}$ is the minimum (maximum) temperature for the target Ising system.
The temperature resolution is defined by $\delta = (T_\text{max} - T_\text{min})/N_\text{conf}$ where $N_\text{conf}$ is the number of samples. 
$\vec{\beta}$ is the discretized inverse temperature defined by
\begin{align}
\text{cl}_N: \beta \to
\vec{\beta} =
\left\{ \begin{array}{ll}
(1,0,\dots,0,0) &\text{for } \beta < 0 \\
(0,1,\dots,0,0) &\text{for } \beta \in [0, \frac{1}{N-2} ) \\
\dots & \\
(0,0,\dots,1,0) &\text{for } \beta \in [\frac{N-3}{N-2}, 1 ) \\
(0,0,\dots,0,1) &\text{for } 1 \leq \beta \\
\end{array} \right.
.
\label{temp_class}
\end{align}
This is called the one-hot representation, and enables us to implement the inverse temperature $\beta$ into the NN directly.
We use index $I$ or $J$ to represent component of the vector $\vec{\beta}$ as already used in .

%%%
%\vspace{-.4cm}
%\subsection{Error function and optimizer}
\vspace{.3cm}
Now let us denote our CNN, explained in \eqref{model} as 
$
F_\text{CNN}^{(F,W)}.
$
We need {\it error function} that measures the difference between the output of the CNN and the correct discretized inverse temperature.
In our case, the task is classification, so we take cross entropy as the error function:
\begin{align}
E ( \vec{\beta}^\text{CNN} , \vec{\beta})
=
- \sum_{I=1}^N 
\beta_I
\log \beta^\text{CNN}_I,
\label{cross_entropy}
\end{align}
where $\vec{\beta}^\text{CNN} = F_\text{CNN}^{(F,W)} ( \{ \sigma_{xy} \}  )$ and $(\{\sigma_{xy} \} , \vec{\beta}) \in \mathcal{T}_L$.

Roughly speaking, the parameters $w = (F_{ij}^a, W_I^m)$ are updated via
$w \leftarrow w - \epsilon \nabla_w E$ with small parameter $\epsilon$.
More precisely speaking, we adopt a sophisticated version of this method called \textit{Adam} \cite{kingma2014adam} implemented in {\tt{TensorFlow (r0.9)}} \cite{abadi2016tensorflow} to achieve better convergence.

%%%
%\vspace{-.4cm}
%\subsection{Summary so far}
%
\vspace{.3cm}
Our neural network, $F_\text{CNN}^{(F,W)}$, learns the optimal parameters $F = \{ F_{ij}^a \}$ and $W = \{W^m_I \}$ in \eqref{weights} through iterating the optimization of the cross entropy \eqref{cross_entropy} between the answer $\vec{\beta}$ and the output $\vec{\beta}_\text{CNN}$ constructed from stochastically chosen data $(\{\sigma_{xy} \}, \vec{\beta}) \in\mathcal{T}_L$.

\vspace{-.3cm}
\begin{align}
&
\left. \begin{array}{ll}
\hline
{\bf Algorithm}
 \\
\hline
\text{{\bf Require:}
CNN $F_\text{CNN}^{(F,W)}$ \eqref{model}; 
an Ising dataset $\mathcal{T}_L$ \eqref{training_set} }
%\\
%\qquad \quad \quad
%\text{
%with fixed $L, N_{\text{conf}}, T_\text{max}, T_\text{min}$
%}
\\
\hline
\text{\tt {\bf for} Num\_of\_iterations {\bf do}} & \\
\text{\tt }
\quad
 \ \ \ \ \text{\tt Choose $(\{ \sigma_{xy} \}, \vec{\beta}) \in \mathcal{T}_L$ randomly}
 & \\
 \text{\tt }
\quad
\ \ \ \ \vec{\beta}^\text{CNN} = F_\text{CNN}^{(F,W)} ( \{\sigma_{xy} \}  )
 & \\
 \text{\tt }
\quad
 \ \ \ \ \text{\tt loss} = E(\vec{\beta}^{\text{CNN}} , \vec{\beta}) \quad \text{\tt by  \eqref{cross_entropy}}
 & \\
 \text{\tt }
\quad 
 \ \ \ \ \text{\tt Update $F, W$ via AdamOptimizer(loss)}
& \\
\text{\tt  {\bf end for}}\\
\hline
\end{array} \right.
\label{alg}
%\notag
\end{align}
%%%
%After the iteration, we check the values for $F$ and $W$ by plotting these components as heat maps.
As we show later, the weight matrix $W$ inheres well approximated critical temperature after 10,000 iterations in \eqref{alg}.

%\newpage
%\vspace{-.6cm}
%\section{Experiments} \label{Result}
%Now, we show our experimental results and how our model \eqref{model} captures the phase transition of the 2D Ising model.
%%%
Here, we prepare $\mathcal{T}_{L}$ \eqref{training_set} by using the Metropolis method with the parameters
\begin{align}
T_\text{min} = 0.1,
\
T_\text{max} = 5.0,
\
N_\text{conf} = 10^4
.
\label{our_training_set}
\end{align}
The max and min values for $T$ mean that $0.2 <\beta<10$, where $\beta$ is the inverse temperature.
Note that the known value form the phase transition is $T_c\sim 2.27$ or $\beta_c \sim 0.44$.
This means that configurations in our training data $\mathcal{T}_{L}$ extend from the ordered phase to the disordered phase. 
In all, we prepare three types of training set, 
\begin{align}
\mathcal{T}_{L=8}
, \
\mathcal{T}_{L=16}
, \
\mathcal{T}_{L=32}.
\end{align}
We apply negative magnetic field weakly to realize the unique ground state at zero temperature.
As a result, almost all configurations at low temperature ($\beta \gg \beta_\text{c} \sim 0.44$) are ordered to $\{ \sigma_{xy} \} = \{ -1, -1, \dots, -1\}$.
%
%%%

%\vspace{-.3cm}
\vspace{.5cm}
%\subsection{No-filter experiment}
Before showing the CNN result, let us try a somewhat primitive experiment: training on NN without the convolution layer, \textit{i.e.} a fully connected NN.
\begin{align}
\left[ \begin{array}{l}
\mathcal{I}=
\{ \sigma_{xy} | \text{ Ising config on }L\times L \text{ lattice.} \}
%\notag 
%\\
% \downarrow
%\
%\left\{ \begin{array}{ll}
%\text{Flatten} & \\
%\end{array} \right.
%\notag 
%\\
%\{ -1 , +1 \}^{L \times L }
%\notag 
\\
\downarrow
\ 
\left\{ \begin{array}{ll}
\text{Flatten} &\\
\text{Fully connected $F$} & \\
\text{Softmax} & \\
\end{array} \right.
%\notag 
\\
\text{[0,1]}^{ \text{Hidden} = 80  }
\\
\downarrow
\ 
\left\{ \begin{array}{ll}
\text{Fully connected $W$} & \\
\text{Softmax} & \\
\end{array} \right.
%\notag 
\\
\text{[0,1]}^{ N = 100  }
= \mathcal{O}
\end{array} \right]
\label{modelFC}
\end{align}
We retain the error function and optimizer, \textit{i.e.} the cross entropy \eqref{cross_entropy} and {{\tt AdamOptimizer}}($%\epsilon = 
10^{-4}$).
We align the heat map for the weight $W$ trained by using $\mathcal{T}_{L=8}$ in right side of \Fig \ref{fig:intro}.
After 10,000 iterations, NN does detected the phase transition.
So this NN is sufficient for detecting ising phase transition, but we cannot answer \textit{why} this NN captures it.
To answer it, we turn to our main target: CNN below.
%

\begin{figure}[t]
\centering
\includegraphics[width=7.0truecm,clip]{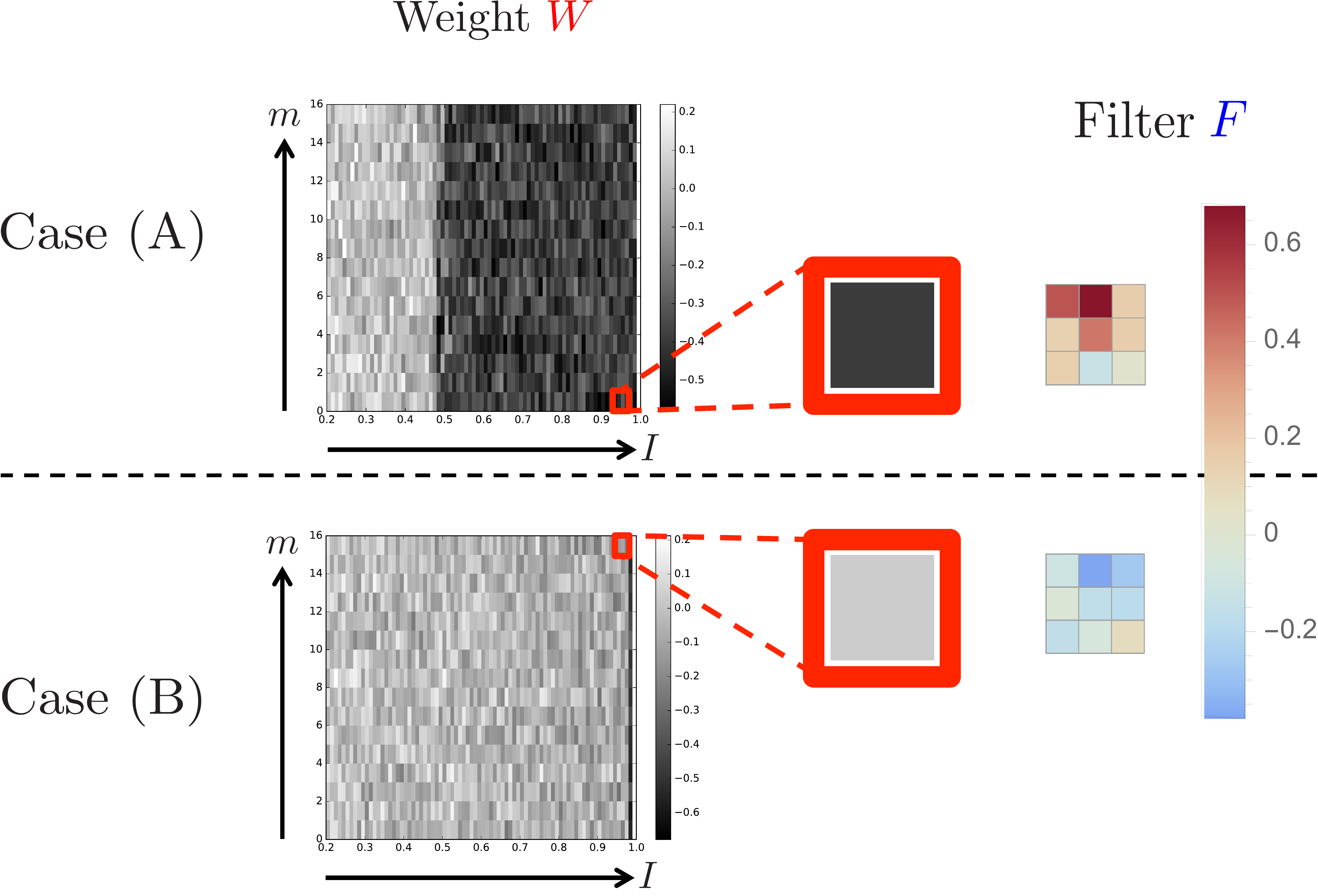} 
\caption{
`(Color online)'
Heat maps of $W^m_I$ and $F_{ij}$ for the CNN with one filter.
In case (A), there always exist two distinct regions (black and gray).
In case (B), there is no such clear decomposition.
\label{fig:test2} }
\end{figure}

%%%
\vspace{.5cm}
%\subsection{One-filter CNN experiment}
Next, we take $N=100, N_f = 3$ and $C =1$ and use the $\mathcal{T}_{L=8}$ training set.
%Just after starting the training, say after 100 iterations, there is no ordered structure as in the no-filter case.
%However, o
Once we increase the number of iterations to 10,000, we get two possible ordered figures.
We denote them case (A) and case (B) respectively as shown in \Fig \ref{fig:test2}.
\\
Case (A) is characterized by $\sum_{ij} F_{ij} >0$, and we can observe two qualitatively different regions in the heat map of the weight $W$, black colored region ($0.48 \lesssim \beta$) and gray colored region ($\beta \lesssim 0.48$).
The boundary is close to the critical temperature $\beta_c \sim 0.44$.
%\footnote{If we take finite size effects into account, the pseudo-critical temperature might be different from the exact value.% $\beta_c \sim 0.44$.}.
%
Case (B) is characterized by $\sum_{ij} F_{ij} <0$, and values in the heat map for $W$ are in gray colored region and almost homogeneous.
We will discuss later the reason why only case (A) displays phase transition.

\begin{figure}[t]
\centering
\includegraphics[width=6.truecm,clip]{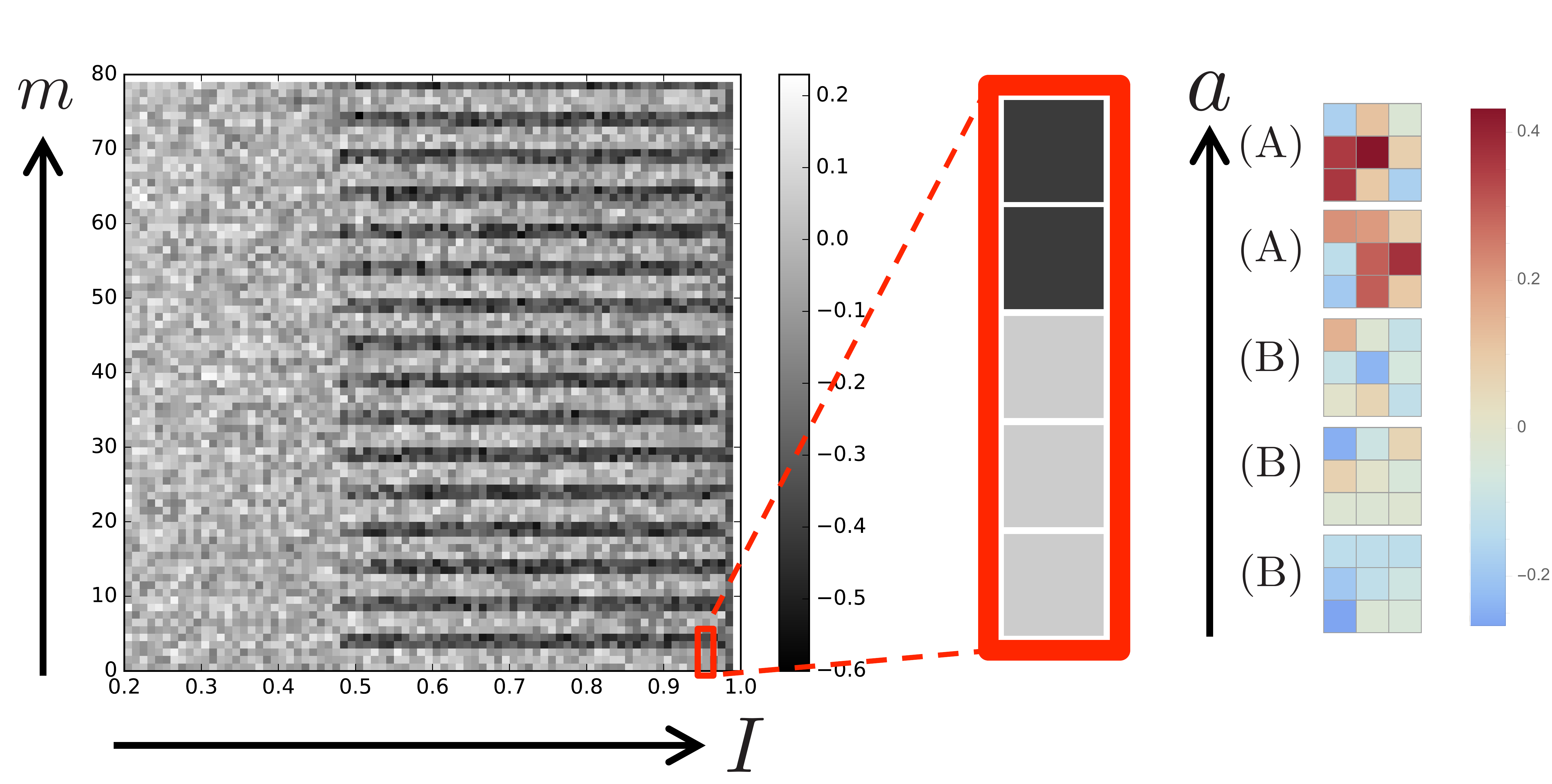} 
\caption{
`(Color online)'
Heat maps of $W^m_I$ and $F_{ij}^a$ with five filters.% 
\label{fig:test3} }
\end{figure}

%%%
%\vspace{-.45cm}
%\subsection{Multi-filter CNN experiment}
%\vspace{-.2cm}
We now turn to the multi-filter case with $N=100, N_f =3, C=5$ and ${L=8}$.
The results for all heat maps after 10,000 iterations are shown in \Fig \ref{fig:test3}.
The stripe structure in the heat map of $W$ corresponds to its values connecting to the convoluted and flatten nodes via five filters, (A), (A), (B), (B), (B) respectively.
Empirically speaking, the number of filters should be large to detect the phase transition because the probability for appearance of (A) increases with increased statistics.
%
%\vspace{-.1cm}
%%%%%

%\newpage

%%%%%
\vspace{.5cm}
%\section{Order parameter based on CNN}\label{Order parameter based on CNN}\label{Order parameter}
From the experiments, we know that our model \eqref{model} seems to discover the phase transition in the given training data for the Ising model.
In order to verify this statement, we would like to extract the critical temperature from our CNN after the training.
As a trial, we fix the parameters of the CNN as follows:
the number of filters, channel and stride are $N_f = 3$, $C=5$ and $s=L/4$ respectively.
The number of classifications is taken as $N=100$ as well as we did in previous section. %
%We mainly use the ReLU activation in \eqref{model}.
%

\begin{figure}[ht]
\centering
\includegraphics[width=7.0truecm,clip]{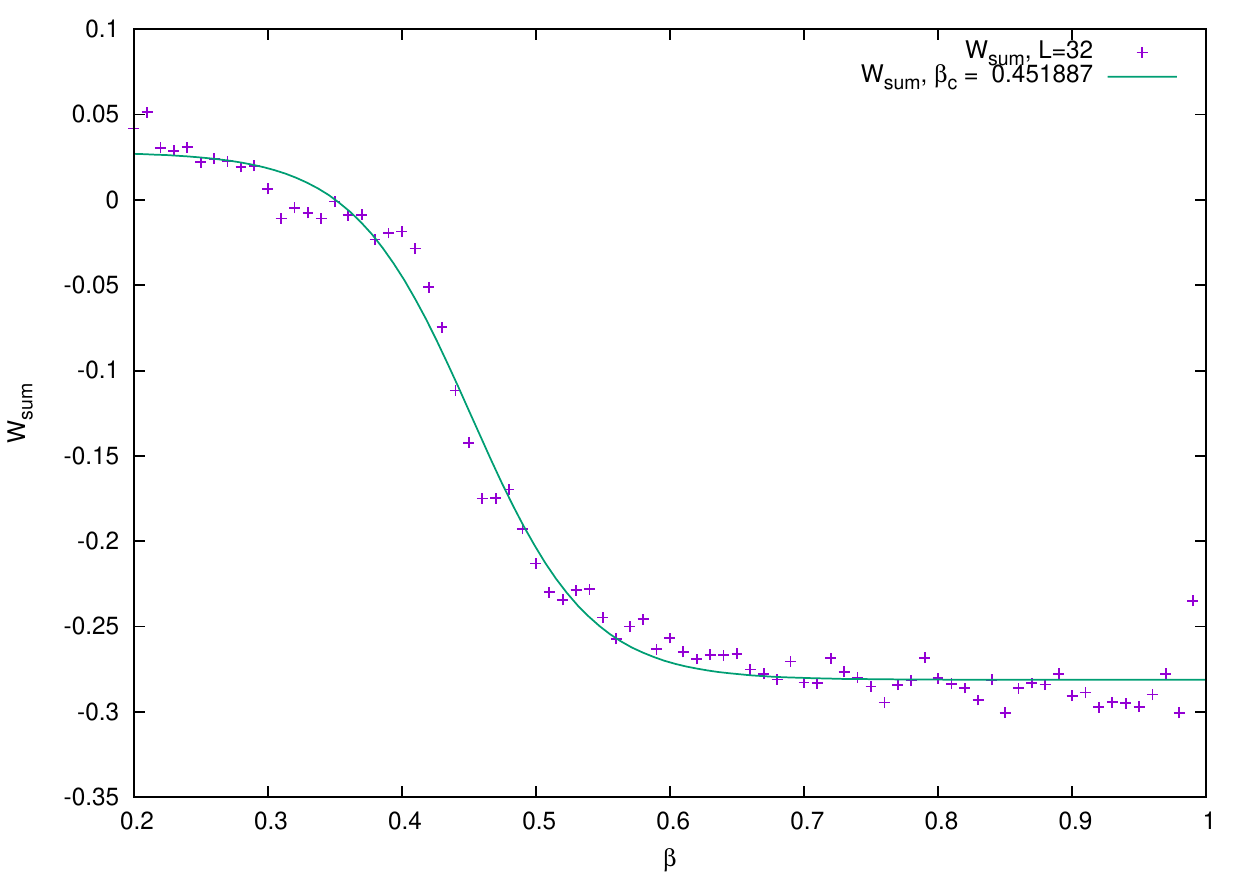} 
\caption{
`(Color online)'
$W_\text{sum}(\beta)$ for $L=32$.
The horizontal axis is the inverse temperature $\beta$ which is translated using \eqref{temp_class}. 
We fit $W_\text{sum}(\beta)$ data by the following smooth function: $a \tanh [ c (\beta - \beta_\text{CNN}) ]  - b$ (Green curve), where $a, b, c, \beta_\text{CNN}$ are fitting parameters, and regard the determined fitting parameter $\beta_\text{CNN}$ as a critical temperature determined by CNN.
\label{fig:result_W_sum_map} }
\end{figure}

First, we plot heat maps for the weight matrix $W$ in CNN trained by $\mathcal{T}_{L=8}, \mathcal{T}_{L=16}, \mathcal{T}_{L=32}$.
For every lattice size, %
we observe a domain-like structure with a boundary around $\beta \sim 0.44$.
However, the heat map does not give the location of the boundary quantitatively.
We propose an order parameter based on the weight matrix $W^m_I$:
%\vspace{-.4cm}
%\subsection{Defining an order parameter}
%
%We define the following quantity,
\begin{align}
W_\text{sum}(\beta) = \sum_m W^m_{I} \text{cl}_N^{I} (\beta), \label{eq:Wsum_def}
\end{align}
%where $ W^m_{I}$ is the weight matrix and $\text{cl}_N^{I} (\beta)$ is the $I$-th component of $\text{cl}_N (\beta)$ as defined in \eqref{temp_class}.
and estimate the critical temperature.
The result is shown in Fig. \ref{fig:result_W_sum_map}.
To quantify the boundary in the heat map for $W$, 
we define critical temperature extracted by CNN $\beta_\text{CNN}$ by fitting $W_\text{sum}(\beta)$ with the following function:
$
\tilde{W}_\text{sum}(\beta)=a\tanh(c(\beta-\beta_\text{CNN}))-b \label{eq:def_beta_cr},
$
where $a, b, c$ and $\beta_\text{CNN}$ are fitting variables and 
$\beta_\text{CNN}$ indicates the location of the jump.
This function is motivated by the magnetization of the Ising model using the mean field approximation.
Table \ref{tab:intro} shows the fit results both for CNN and FC.
Our results show that $\beta_\text{CNN}$ matches the critical temperature to 2 -- 8 \% accuracy.
Compared to it, $\beta_{FC}$ shows less accuracy.
%
\if0
Let us comment here about what happens if we change the activation function for the convolution layer in \eqref{model} from ReLU to ELU.
In that case, no significant phase transition-like signal is obtained through $W_\text{sum}(\beta)$.
We cannot obtain $\beta_\text{CNN}$ by fitting because of the absence of the gap-like structure.
The following results are obtained by using the ReLU.

% 
\begin{table}[htb]
\begin{tabular}{c||cc|cc} 
$L$ & $\beta_\text{CNN}$ %& $\chi^2/\text{d.o.f.}$
& $T_c$ accuracy
& $\beta_\text{FC}$
& $T_c$ accuracy
\\
\hline
8
& 0.478915%& 0.000267911
& 8.12 \%
& Hoge
& Wow
\\
16& 0.448562%& 0.000273062
& 1.91 \%
& Hoge
& Wow
\\
32& 0.451887%& 0.000180452
& 2.63 \%
& Hoge
& Wow
\end{tabular}
\caption{
Critical temperature defined by the fit (\ref{eq:def_beta_cr}).
$\mathcal{O}$(1) \% error is expected because the resolution of the inverse temperature is $1/N = 1/100$.
\label{tab:beta_critical_summary}}
\end{table}
\fi

\vspace{.5cm}
%\section{Summary and Discussion}\label{Discussion}
%
Let us conclude this \this.
We have designed simple neural networks to study correlation between configuration of the 2D Ising model and inverse temperature, and we have trained them by SGD method implemented by {\tt TensorFlow}.
We have found that the weight $W$ in neural networks captures a feature of phase transition of the 2D Ising model, and defined a new order parameter $W_\text{sum}(\beta)$ in \eqref{eq:Wsum_def} via trained neural networks and have found that it can provide the value of  critical inverse temperature.

Why are our neural networks able to find a feature of phase transition?
There is an intuitive explanation thanks to CNN experiments.
The filter with $N_f =3$ in case (A) has a typical average around $0.1\sim 0.2$. 
This is close to the convolution with filter $F_{ij} = 1/N_f^2$ which is equivalent to a real space renormalization group transformation, and the filters reflect local magnetization which is related to a typical order parameter and it enables CNN to detect the phase transition.
As an analog of this, FC NN might realize the real space renormalization group transformation in inside.
  
Our NN model has potential to investigate other statistical models.
For example, it was reported that CNNs can distinguish phases of matters, topological phases in $\mathbb{Z}_2$ gauge theories \cite{carrasquilla2016machine}, phases in the Hubbard model \cite{ch2016machine} and Potts model \cite{potts}.
It is interesting to apply our design of neural networks to these problems and see whether the NN can discover nontrivial phases automatically, as we did in this \this.

\vspace{.5cm}
%%%%%%%%%%%%%%%%%%%%%%%%%%%%
\begin{acknowledgment}
We would like to thank to K. Doya, K. Hashimoto, T. Hatsuda, Y. Hidaka, M. Hongo, B. H. Kim, J. Miller, S. Nagataki, N. Ogawa,  M. Taki and Y. Yokokura for constructive comments and warm encouragement.
We also thank to D. Zaslavsky for careful reading this manuscript.
The work of Akinori Tanaka was supported in part by the RIKEN iTHES Project.
The work of Akio Tomiya was supported in part by NSFC under grant no. 11535012.
\end{acknowledgment}
%%%%%%%%%%%%%%%%%%%%%%%%%%%%
\newcommand{\newblock}{} % added to put arxiv
\bibliographystyle{jpsj}
\bibliography{Ref}
%%%%%%

\end{document}